\documentclass{article}

\usepackage{arxiv}

\usepackage[utf8]{inputenc} 
\usepackage[T1]{fontenc}    
\usepackage{hyperref}       
\usepackage{url}            
\usepackage{booktabs}       
\usepackage{amsfonts}       
\usepackage{nicefrac}       
\usepackage{microtype}      
\usepackage{lipsum}
\usepackage{graphicx}
\graphicspath{ {./images/} }
\usepackage{amsmath} 
\usepackage{multirow}

\title{SciRGC: Multi-Granularity Citation Recommendation and Citation Sentence Preference Alignment}

\author{
  Xiangyu Li \\
  Nanjing University of Posts and Telecommunications \\
  Nanjing, China \\
  \texttt{32804131lxy@gmail.com} \\
  \And
  Jingqiang Chen \\
  Nanjing University of Posts and Telecommunications \\
  Nanjing, China \\
  \texttt{cjq@njupt.edu.cn}
}

\begin{document}
\maketitle
\begin{abstract}
Citations are crucial in scientific research articles as they highlight the connection between the current study and prior work. However, this process is often time-consuming for researchers. In this study, we propose the SciRGC framework, which aims to automatically recommend citation articles and generate citation sentences for citation locations within articles. The framework addresses two key challenges in academic citation generation: 1) how to accurately identify the author's citation intent and find relevant citation papers, and 2) how to generate high-quality citation sentences that align with human preferences. We enhance citation recommendation accuracy in the citation article recommendation module by incorporating citation networks and sentiment intent, and generate reasoning-based citation sentences in the citation sentence generation module by using the original article abstract, local context, citation intent, and recommended articles as inputs. Additionally, we propose a new evaluation metric to fairly assess the quality of generated citation sentences. Through comparisons with baseline models and ablation experiments, the SciRGC framework not only improves the accuracy and relevance of citation recommendations but also ensures the appropriateness of the generated citation sentences in context, providing a valuable tool for interdisciplinary researchers.
\end{abstract}


\section{Introduction}

Citation sentence is a sentence that starts from the content of a cited paper, identifies problems, contradictions, gaps, or similarities related to the current state of the art, and integrates them systematically, strategically, and critically\cite{1} to help readers understand the contributions of the cited paper versus those of the cited paper in terms of theories, methods, or discoveries\cite{2}. When writing scientific articles, researchers need to find relevant results from a large body of previous work and cite them appropriately at the right place in the article, which is a time-consuming and labor-intensive process\cite{3}, and especially challenging for new scholars\cite{4}. Automating the process of quickly targeting and summarizing literature relevant to the context of the current paper for citation by understanding the author's citation intent, ensuring accuracy, and generating fluent citations is a challenging task.

\begin{figure}[ht]
    \centering
    \includegraphics[width=0.75\linewidth]{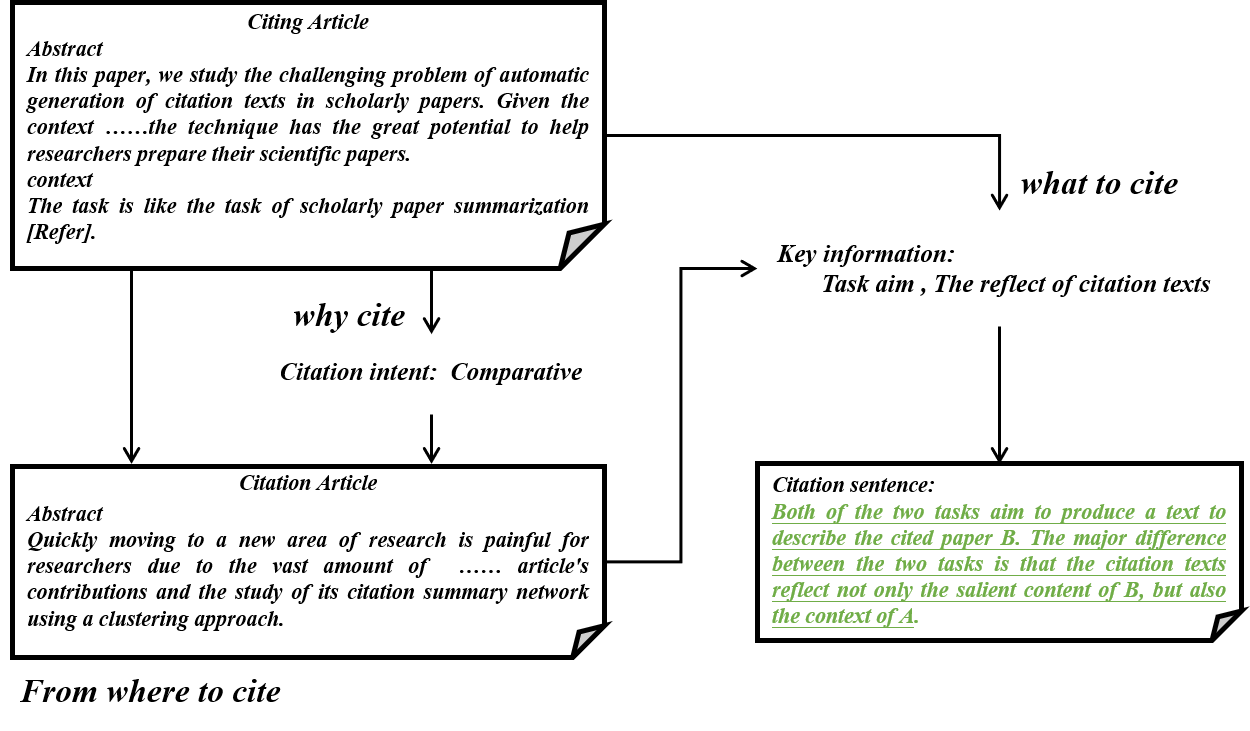}
    \caption{In the process of citation generation, it is first necessary to infer the citation intent, then to find the citation article, and finally to generate the reasoning-based citation sentence.}
    \label{fig introduction}
\end{figure}

Previous research on citation of scientific papers has covered several directions, such as citation functional classification\cite{5, 6, 7, 8}, citation recommendation\cite{9, 10, 11}, citation generation and article abstraction\cite{12, 13, 14, 15, 16, 17}, etc.  However, the correlation among the controllability of citation intent for citation recommendation and citation sentence generation has not been fully explored.

Our research focuses on how to make full use of the various characteristics of citations quickly target relevant papers and then generate citation sentences given the citing location in the paper. An example of this process is shown in Figure  \ref{fig introduction}. This example is from \cite{14}. We first infer the author's citation intent based on the abstract of the article with the local context, then find a suitable article to cite based on this intent, and finally generate a suitable citation sentence by reasoning about what the author wants to cite. In this study, we elaborate on the challenges and solutions at each stage of the process and propose the SciRGC framework consisting of a citation articles recommendation module and a citation sentence generation module.

For citation articles recommendation, the input is the query text without citations, while the output is the citation-worthy article\cite{18}. Traditional citation recommendation methods fall into two categories: global citation recommendation and local citation recommendation. The former focuses on the information of the entire article and aims to recommend citable article as a whole\cite{19, 20}, usually using the title and abstract as the query text; And the latter considers the literature's in-text specific location and its context\cite {21, 22}, and uses the context as the feature text input. Our approach considers both global and local information with the recall stage and the reranking stage. In the recall stage, we use an encoder model to capture local information and a collaborative filtering algorithm using a citation network to obtain a global perspective, which are combined to obtain a preliminary list of relevant articles. In the reranking stage, we improve the accuracy and controllability of the recommendation by recognizing the authors' citation intentions, based on which the articles list is adjusted to achieve a better balance.

In citation sentence generation, the input consists of the abstract of the original paper, the local context at the citation position, the citation intent, and the abstract of the recommended cited paper, while the output is an inference-driven citation sentence.  This paper proposes an inference-driven citation generation method. The approach first leverages a large language model (LLM) to extract key information for constructing a Chain-of-Thought (CoT) reasoning process. Then, the model undergoes instruction fine-tuning using the LoRA strategy \cite{23} and is further refined through Direct Preference Optimization (DPO) \cite{47} to align outputs with human preferences, ultimately generating high-quality, inference-based citation sentences.

This study conducts a systematic experimental validation on two key tasks: citation recommendation and citation generation.  For the citation recommendation task, experiments are conducted on four academic literature datasets—ACL-200, FullTextPeerRead, RefSeer, and arXiv—using Mean Reciprocal Rank (MRR) and Recall at K (R@K) as the core evaluation metrics. The results demonstrate that the proposed recommendation method consistently outperforms existing approaches across all benchmark datasets.  For the citation generation task, experiments are conducted on the Citation Generation (CG) dataset \cite{14}. Addressing the systematic bias of traditional automatic evaluation metrics, this study introduces an innovative multidimensional citation generation quality evaluation framework. Human evaluation confirms the validity of this framework, and under this evaluation system, the proposed generation model achieves superior performance compared to mainstream large language models. The contributions of this study can be summarized as follows:

\begin{itemize}
\item For the citation article recommendation stage, this paper proposes using a citation network and citation intent to enhance the accuracy and controllability of recommended cited papers, with significant improvements observed.
\item For citation sentence generation, this paper proposes a citation fine-tuning Chain-of-Thought method for large language models, which greatly improves the quality of generated citation text.
\item This paper also introduces a new evaluation metric for citation sentences, effectively addressing the limitations of traditional machine evaluation metrics.
\end{itemize}

\section{Related Work}

The citation recommendation task focuses on a two-stage strategy of prefetching and sorting. \cite{21} nonparametric probabilistic model laid the foundation for early local citation recommendation, while recent studies prefer embedding techniques to match precisely by cosine or Euclidean distance. \cite{28} combines BERT\cite{29} with Graph Convolutional Networks (GCN)\cite{30} to dig deeper into the paper's association, but this method is limited by the high computational cost of GCN. \cite{31} innovate transformer encoding, introduce paragraph attributes, and optimize sorting by scibert\cite{32} to improve citation scores, where the prefetching reordering strategy improves citation scores.  From the point of view of the dataset, supervised recommendation methods seek for massive annotation, ACL-200\cite{11}, FullTextPeerRead\cite{28} and RefSeer\cite{11} are limited in size, RefSeer data is obsolete, arXiv\cite{31} to S2ORC\cite{33} huge corpus, to construct a new citation recommendation database, to meet the needs of the academic community.

In the field of citation generation, scholars have explored a variety of innovative ways to improve the quality and efficiency of generation.\cite{14} identify explicit and implicit citations, and the PTGEN-Cross model relies on the pointer network to replicate the abstract vocabulary and vividly display the citation dynamics.\cite{17} introduce citation graphs and intents to deepen the information hierarchy and refine the generation process.\cite{15} Integrate document summaries and quotations, apply GPT-2\cite{27} model for generation, and produce natural and fluent citations with rich connotations.\cite{34} propose a comparative graph-based summarization (CGSUM) method,  generates summaries with comparative analysis nature using the citations as a guideline.\cite{26} is dedicated to controlled citation generation, utilizing models such as BART\cite{35} to make the citation generation process more flexible and able to adjust the generation results to specific needs.

\section{Methodology}
\begin{figure*}[ht]
\centering
\includegraphics[width=1\textwidth]{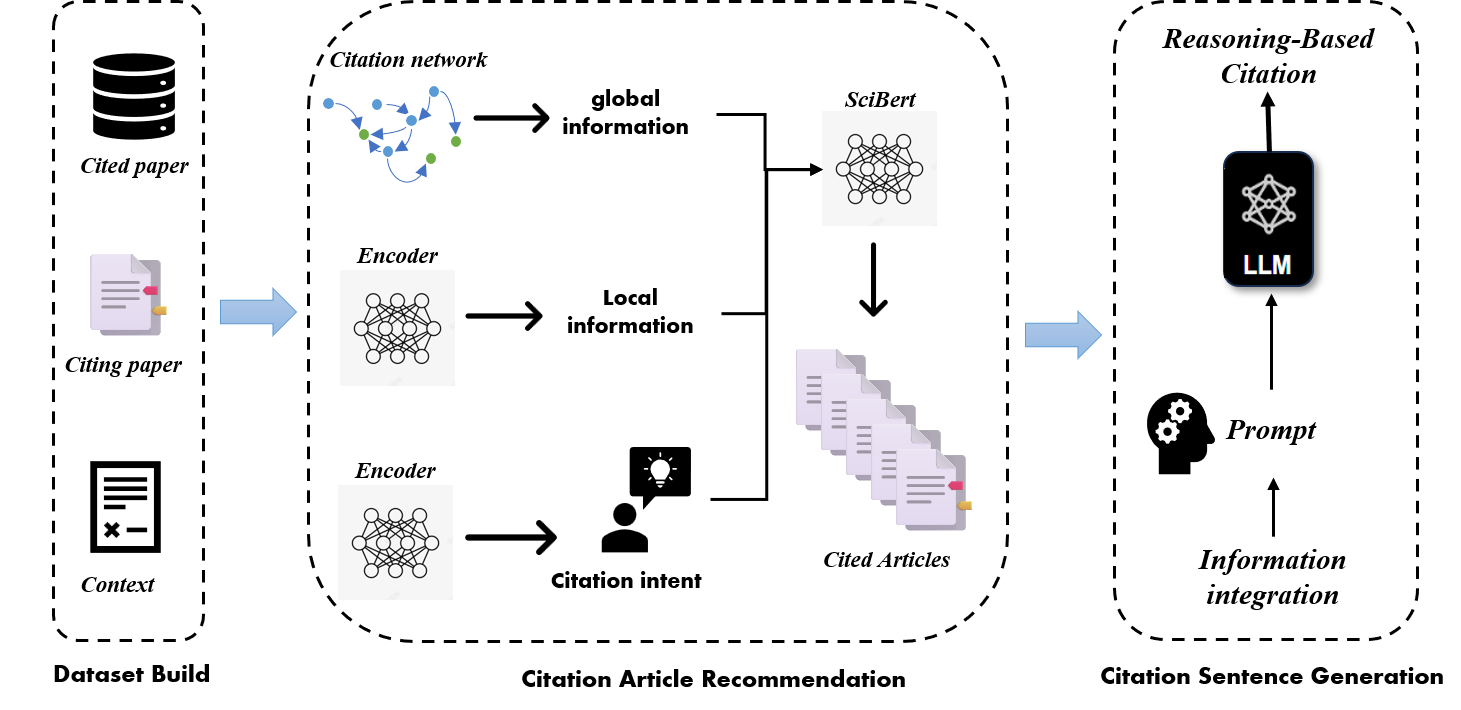}
\caption{Process for implementing citation recommendation and generation in the SciRGC framework}
\label{fig 1}
\end{figure*}

Our proposed framework aims to find cited articles that match the citation intent of the moment based on the abstracts and local contexts of the citing paper and then combine the information to generate reasoning-based citation sentences. As shown in Figure \ref{fig 1}, the framework consists of two parts: citation article recommendation and reasoning-based citation sentence generation. 

\subsection{Citation Article Recommendation}
Building upon \cite{31}, this study optimizes two stages of the citation recommendation system: the fast recall model and the slower re-ranking model. The key difference lies in the recall stage, where both local and global information are integrated, while the re-ranking stage incorporates citation intent features to enhance the recommendation results.

\subsubsection{Recall model}
In the recall phase, we blend local and global information by proposing two collaborative filtering algorithms based on citation networks to refine and utilize the global correlations among documents. 

\begin{figure}[ht]
\centering
\includegraphics[width=0.5\columnwidth]{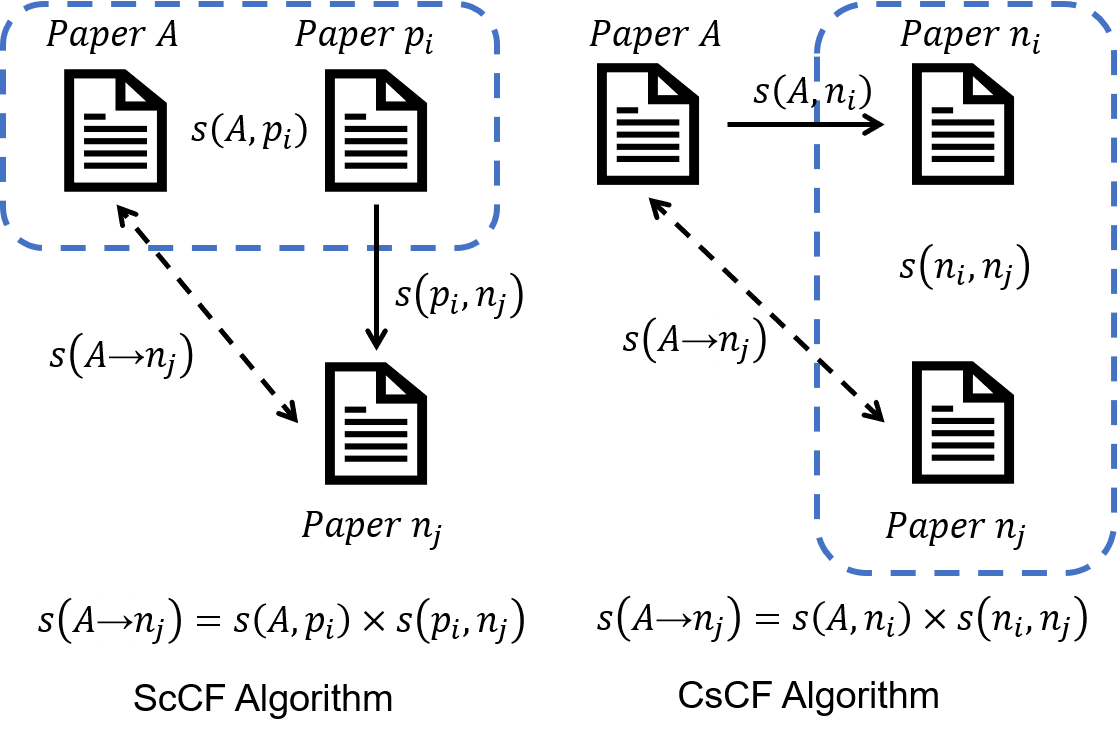} 
\caption{Two collaborative filtering algorithms, dashed boxes represent two papers that are similar, and single arrows indicate citation of the paper.}
\label{collaborative_filtering}
\end{figure}

To calculate the global recall score, this paper constructs a citation network, where papers are represented as nodes, and edges indicate direct citation relationships. Based on this network, and inspired by previous research \cite{36}, we design two collaborative filtering algorithms as shown in fig \ref{collaborative_filtering}. The core assumption is that if two papers share a large number of the same citations, they are likely to belong to the same research field and exhibit high similarity. Conversely, if they share fewer citations, their relevance is weaker.

The collaborative filtering algorithm based on the similarity of citing papers (ScCF) is characterized by its use of structured citation information and semantic similarity to predict the citation behavior of papers. It infers the papers that an article may cite by analyzing the similarity relationship between articles in the citation network. For example, if article A is similar with article B in the citation network and article A cites article C, article B can also cite article C. This approach emphasizes the semantic and citation relationships between articles, which are inferred with the help of structured data.

In contrast, the Collaborative filtering based on the similarity of cited papers (CsCF) focuses more on the similarity between cited literature. It also uses the similarity relationships of articles in the citation network to infer the citation behavior of papers but focuses on analyzing whether other articles cited by an article are similar. For example, if article A cites articles B, C, and D, and articles B, C, and D are similar to article E, it is hypothesized that article A may also cite article E. Instead of focusing directly on the similarity of the entire article, this approach focuses on the association between the cited literature.

\begin{figure}[ht]
\centering
\includegraphics[width=0.5\columnwidth]{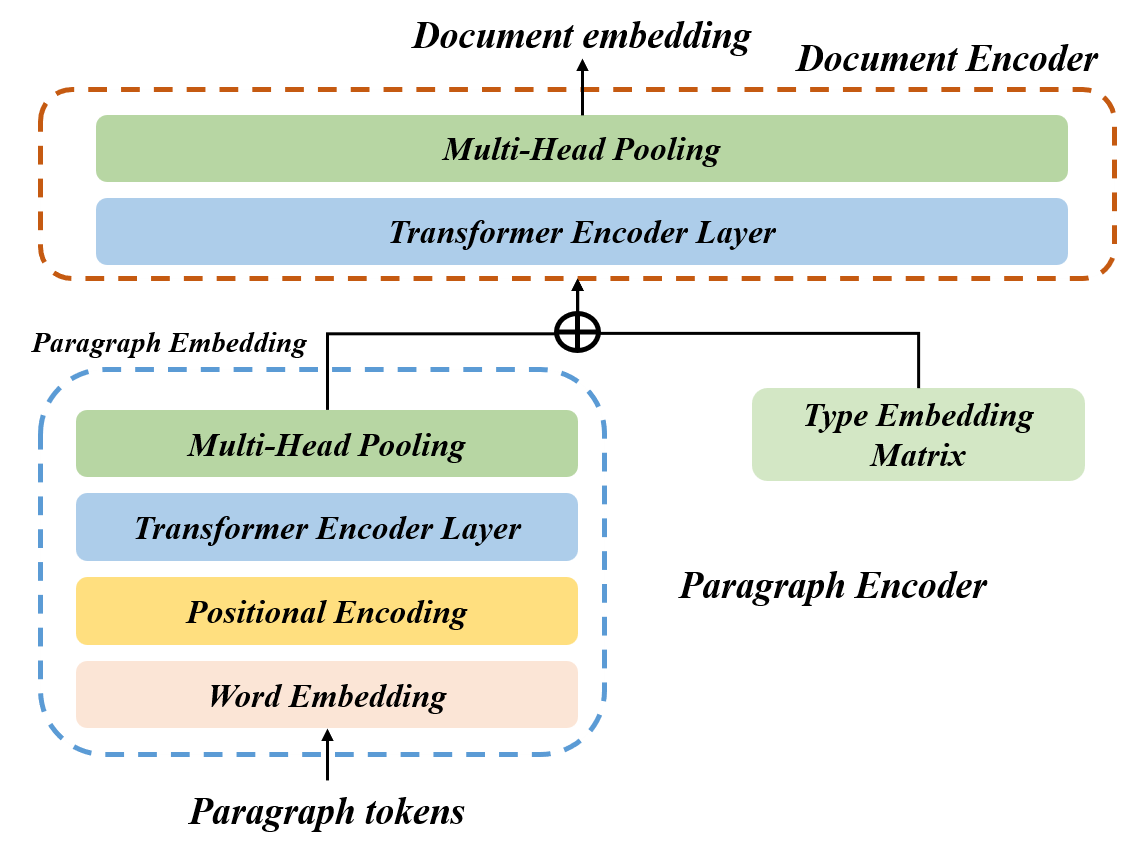} 
\caption{The Encoder proposed in the Recall phase consists of a paragraph encoder and a document encoder, where different weights are assigned to different paragraphs by adding paragraph types in the document encoder section.}
\label{recall_model}
\end{figure}

The encoder for computing the local recall score is shown in Fig \ref{recall_model}. It consists of two parts: the paragraph encoder and the document encoder.

For each paragraph $ p_i $ in the document, the paragraph encoder takes a sequence of $n_i$ tokens $p_i = [w_1, ..., w_{n_i}]$ as input and outputs the paragraph embedding $ e_{p_i} $. The paragraph encoder uses positional coding and encodes the contextual information through a transformer coding layer. To obtain a fixed-size paragraph embedding $ e_p $, the paragraph encoder uses a multi-head pooling layer to process the output of the transformer encoder. For each header $ j$, the value vector $v_j^k $ and the attention score $ \hat{a}_j^k $ are computed:

\begin{equation}
v_j^k = \text{Linear}_j^v (x_k), \quad a_j^k = \text{Linear}_j^a(x_k)
\end{equation}

\begin{equation}
\hat{a}_j^k = \frac{\exp(a_j^k)}{\sum_{m=1}^{n_{token}} \exp(a_j^m)} 
\end{equation}

The weighted value vector $ \hat{v}_j $is the weighted sum of all the value vectors. Activated by ReLU and linearly transformed, the final paragraph embedding $ e_p $ is the combination of the weighted value vectors of all heads:
\begin{equation}
e_p = \text{Linear}_p(\text{ReLU}(\text{Concat}(\hat{v}_1, ... , \hat{v}_{n_{head}}))) 
\end{equation}

After obtaining the document encoding $ p_i $, we combine three types of paragraph representations (title, abstract, and local context) with their corresponding paragraph embeddings. By adding paragraph type embeddings, we generate type-aware paragraph embeddings. These type-aware paragraph embeddings are then fed into the transformer encoder layer, and the final document embedding is obtained through a multi-head pooling layer.

The final ranking of the recalled candidate documents is performed by using a nearest neighbor search based on the similarity scores obtained from weight fusion:

\begin{equation}
S=s_{\text{encoder}} \ast w_1 + s_{\text{CF}} \ast w_2
\end{equation}

where $s_{\text{encoder}}$ is the similarity score obtained by the Encoder part by identifying the $K$ nearest document embeddings embedded with the query based on cosine similarity, and $s_{\text{CF}}$ is the similarity score obtained by collaborative filtering of the paper. $w_1$ and $w_2$ are the weights of the fusion of the two model scores respectively.

\subsubsection{Reranker model}

We incorporate citation intent features in our reordering model to achieve a more accurate match between cited and retrieved candidate documents. We developed a citation intent recognition model based on the method of \cite{17} to identify citation purposes. For details on the definition of citation intent, model parameters, and analysis of results, please refer to the Appendix.

After obtaining the citation intent, we take the citation intent, the abstract of the cited document, the local context of the citation location, and the abstract of the candidate document as inputs, and calculate the relevance score of each candidate document as the model output. Specifically, we utilize SciBERT to encode the input text and extract vector representations of the “[CLS]” tags. These vectors are then fed into a feed-forward neural network and a Sigmoid function is used to generate a similarity score between 0 and 1, which characterizes the degree of fit between the candidate document and the citation location. 

\subsection{Reasoning-Based Citation Sentence Generation}

\begin{figure*}[ht]
\centering
\includegraphics[width=1\textwidth]{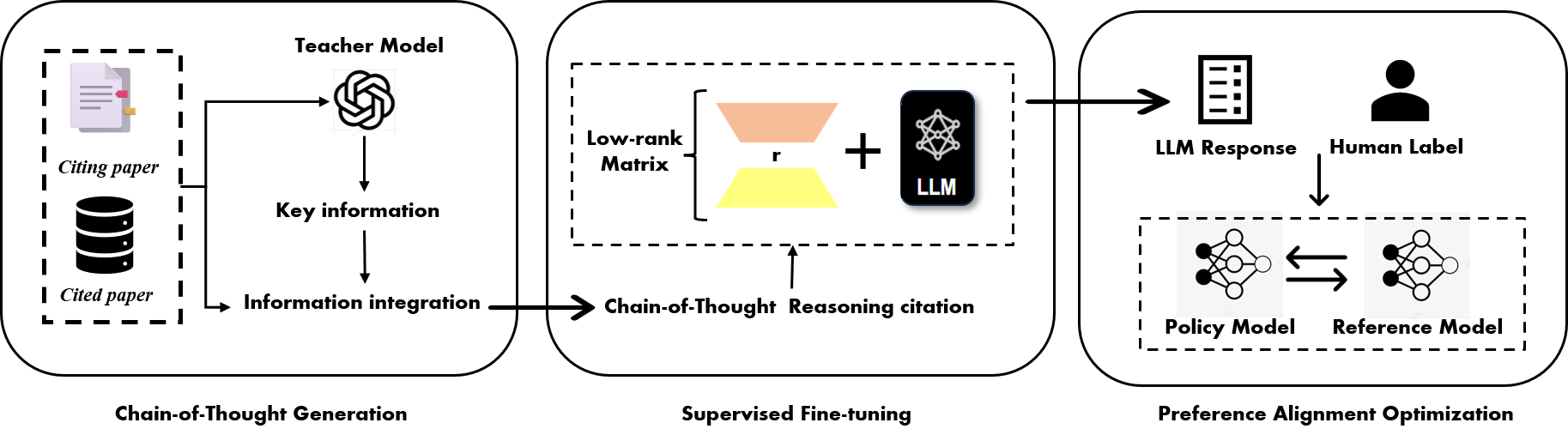}
\caption{The three-phase citation generation framework proposed in this paper}
\label{fig4}
\end{figure*}

The input for the citation generation phase primarily consists of the abstract of the cited paper, the context around the citation position, the abstract of the cited paper obtained through citation recommendations, and the citation intent. The core task is to generate citation text that aligns with the contextual content based on the input. 

This paper proposes a three-phase citation generation method, as shown in Figure \ref{fig4}. The method integrates three strategies: Chain-of-Thought (CoT), Lightweight Supervised Fine-Tuning (SFT), and Direct Preference Optimization (DPO), constructing an efficient and resource-light citation generation framework. The aim of this approach is to enhance the academic quality, accuracy, and information density of the citation text, while effectively reducing the demand for computational resources.

\subsubsection{Chain-of-Thought Generation}
Although large-scale language models (LLMs) have demonstrated exceptional capabilities in natural language processing (NLP), providing a foundation for addressing this challenge, direct fine-tuning may not be effective for compact, information-dense citation texts, as it cannot meet the stringent requirements of academic writing in terms of high information density and precision. Existing methods for addressing the limitations of complex reasoning generation \cite{38} typically involve integrating a small number of chain-of-thought (CoT) reasoning examples into large language models \cite{39}, or using prompts \cite{40} to stimulate the model's complex reasoning potential and guide its progressive thinking. 

 The core idea behind the generation of reasoning data in this paper is to generate the reasoning process for citation generation using a large teacher model, and then use these reasoning results to fine-tune a smaller student model, enabling it to acquire reasoning capabilities \cite{41}. This approach not only effectively reduces computational costs but also maintains high reasoning performance and output quality. The implementation of this approach is as follows: First, GPT-4 \cite{42} is employed for zero-shot learning \cite{43, 40} to extract key information such as themes and keywords from the cited and citing papers. Then, this key information is combined with citation sentences to construct chain-of-thought data, which is concatenated with citation sentences to produce ground truth citation sentences for fine-tuning smaller language models, incorporating the reasoning process.

\subsubsection{Supervised Fine-tuning}
In the citation generation process, model fine-tuning is a key step in improving the quality of the generated text. In this paper, we use the dataset from \cite{14} for fine-tuning. Given that most of the information in large pre-trained language models (LLMs) is concentrated in lower intrinsic dimensions \cite{23}, and that directly fine-tuning LLMs requires significant computational resources and time, we employ a lightweight fine-tuning strategy—Low-Rank Adaptation (LoRA) \cite{44, 45}—to achieve efficient model fine-tuning. LoRA works by freezing most of the pre-trained model parameters and introducing trainable low-rank decomposition matrices into each layer of the Transformer model. This allows fine-tuning with only a small number of parameters, achieving performance comparable to full parameter fine-tuning without altering the original weight matrices. This method not only preserves the model's reasoning ability but also significantly reduces training costs and hardware requirements while effectively integrating additional information.

The optimization objective of LoRA can be expressed as:

\begin{equation}
\min_{\mathbf{A}, \mathbf{B}} \mathcal{L}(f(\mathbf{W} + \mathbf{A} \mathbf{B}), \mathcal{D})
\end{equation}

where $\mathbf{W}$ is the original weight matrix of the pre-trained model, $\mathbf{A}$ and $\mathbf{B}$ are the trainable rank decomposition matrices, $\mathcal{L}$ is the loss function, $f$ is the model function, and $\mathcal{D}$ is the training dataset. With LoRA, we can use only one-tenth of the original LLM parameters to complete the training process \cite{23}.

\subsubsection{Preference Alignment Optimization}

The quality of citation generation is not only reflected in the fit with the context and the accuracy of the information but also in the academic nature and readability of the text. To further improve the quality of generated citations, this paper introduces the DPO (Direct Preference Optimization) \cite{47}strategy. DPO optimizes the generated citations by maximizing the log probability ratio between high-quality and low-quality citation texts, making them more in line with academic writing standards and closer to human preferences.

Specifically, the citation texts generated by SFT serve as the lower-preference (loser) samples, while the high-quality citation texts manually annotated serve as the higher-preference (winner) samples. DPO performs comparative optimization to maximize the log probability ratio of the higher-preference text relative to the lower-preference text, with the optimization objective function being:

\begin{equation}
L_{DPO} = -E_{(X, Y_w, Y_l) \sim \mathcal{D}} \log \sigma(r_\theta(X, Y_w) - r_\theta(X, Y_l))
\end{equation}

Here, $ Y_w $ and $ Y_l $ represent the higher-preference and lower-preference citation texts, respectively, $ \sigma $ is the sigmoid function, $ r_\theta $ is the score of the reward model, and $ \mathcal{D} $ is the training dataset. By optimizing this objective function, DPO makes the generated citations more in line with human preferences, thereby enhancing the academic nature and readability of the text.

\section{Evaluation Criteria}
The construction of academic citations faces dual challenges: first, achieving a balance among multiple requirements (such as information accuracy and contextual coherence); second, the lack of systematic functional evaluation standards. Based on an interdisciplinary analysis of citation examples in computer science, biomedicine, and social sciences, we identify four core dimensions of high-quality citation construction:

\subsection{Core Dimensions of Citation Construction}

\subsubsection{Purpose-driven Articulation}
An effective citation should actively strengthen the argumentative logic of a paper rather than serve as a decorative element. Corpus analysis reveals that high-quality citations establish explicit connections with the research question of the host paper, facilitating a functional integration of external evidence and internal reasoning. For instance, in methodological justification, expert authors tend to cite literature that demonstrates methodological innovation rather than merely listing similar studies.

\subsubsection{Semantic Accuracy}
Citations must accurately convey the core content of the original research, adhering to the principle of fidelity. This ensures that the cited information remains precise and unaltered, preserving the integrity of academic discourse and enabling readers to trace the origins of ideas. This dimension requires an isomorphic mapping between the original intent of the cited work and its presentation in a new context. In this study, we assess accuracy through three sub-criteria: conceptual precision (preservation of key definitions), methodological consistency (appropriateness of technical descriptions), and cognitive alignment (correct interpretation of research findings).
\begin{table*}[ht]
\centering
\renewcommand{\arraystretch}{1}
\begin{tabular}{c|cccccccc}
\hline
Dataset                           & Model              & MRR            & R@10           & R@100          & R@200          & R@500          & R@1000         & R@2000         \\ \hline
\multirow{4}{*}{ACL-200}          & BM25               & 0.138          & 0.263          & 0.52           & 0 .604         & 0.712          & 0.791          & 0.859          \\
                                  & Sent2vec           & 0.066          & 0.127          & 0.323          & 0.407          & 0.533          & 0.64           & 0.742          \\
                                  & HAtten             & 0.154          & 0.269          & 0.577          & 0.679          & 0.798          & 0.865          & 0.924          \\
                                  & Atten-CF & \textbf{0.186} & \textbf{0.312} & \textbf{0.605} & \textbf{0.702} & \textbf{0.810} & \textbf{0.873} & \textbf{0.926} \\ \hline
\multirow{4}{*}{FullTextPeerRead} & BM25               & 0.185          & 0.328          & 0.609          & 0.694          & 0.802          & 0.877          & 0.95           \\
                                  & Sent2vec           & 0.121          & 0.215          & 0.462          & 0.561          & 0.694          & 0.794          & 0.898          \\
                                  & HAtten             & 0.178          & 0.31           & 0.653          & 0.76           & 0.875          & 0.941          & 0.982          \\
                                  & Atten-CF & \textbf{0.180} & \textbf{0.329} & \textbf{0.666} & \textbf{0.770} & \textbf{0.881} & \textbf{0.944} & \textbf{0.984} \\ \hline
\multirow{4}{*}{RefSeer}          & BM25               & 0.099          & 0.189          & 0.398          & 0.468          & 0.561          & 0.631          & 0.697          \\
                                  & Sent2vec           & 0.061          & 0.111          & 0.249          & 0.306          & 0.389          & 0.458          & 0.529          \\
                                  & HAtten             & 0.121          & 0.225          & 0.498          & 0.596          & 0.725          & 0.798          & 0.87           \\
                                  & Atten-CF & \textbf{0.145} & \textbf{0.249} & \textbf{0.517} & \textbf{0.619} & \textbf{0.734} & \textbf{0.807} & \textbf{0.878} \\ \hline
\multirow{4}{*}{arXiv}            & BM25               & 0.118          & 0.222          & 0.451          & 0.529          & 0.629          & 0.7            & 0.763          \\
                                  & Sent2vec           & 0.072          & 0.131          & 0.287          & 0.347          & 0.435          & 0.501          & 0.571          \\
                                  & HAtten             & 0.124          & 0.241          & 0.527          & 0.619          & 0.734          & 0.809          & 0.871          \\
                                  & Atten-CF & \textbf{0.128} & \textbf{0.255} & \textbf{0.536} & \textbf{0.626} & \textbf{0.751} & \textbf{0.820} & \textbf{0.881} \\ \hline
\end{tabular}
\caption{Experimental Results of Citation Recommendation Recall Modeling}
\label{Recall Results}
\end{table*}

\begin{table*}[ht]
\centering
\renewcommand{\arraystretch}{1}
\begin{tabular}{c|cccccccc}
\hline
\multirow{2}{*}{Model} & \multicolumn{2}{c}{\textbf{ACL-200}} & \multicolumn{2}{c}{\textbf{FullTextPeerRead}} & \multicolumn{2}{c}{\textbf{RefSeer}} & \multicolumn{2}{c}{\textbf{arXiv}}\\  
                       & MRR               & R@10             & MRR                   & R@10                  & MRR               & R@10             & MRR              & R@10            \\ \hline
NCN                    & -                 & -                & -                     & -                     & 0.267             & 0.291            & -                & -               \\
DualCon                & 0.335             & 0.647            & -                     & -                     & 0.206             & 0.406            & -                & -               \\
DualEnh                & 0.366             & 0.703            & -                     & -                     & 0.28              & 0.534            & -                & -               \\
BERT-GCN               & -                 & -                & 0.418                 & 0.529                 & -                 & -                & -                & -               \\
BERT                   & 0.482             & 0.736            & 0.458                 & 0.706                 & 0.309             & 0.535            & 0.226            & 0.399           \\
SciBERT                & 0.531             & 0.779            & 0.536                 & 0.773                 & 0.380             & 0.623            & 0.278            & 0.475           \\
SciBERT-Intent         & \textbf{0.552}    & \textbf{0.797}   & \textbf{0.554}        & \textbf{0.784}        & \textbf{0.400}    & \textbf{0.650}   & \textbf{0.291}   & \textbf{0.496}  \\ \hline
\end{tabular}
\caption{Experimental Results of Citation Recommendation Reranking Modeling}
\label{Reranking Results}
\end{table*}

\subsubsection{Contextual Fit}
Citations should seamlessly integrate with the surrounding discussion, maintaining coherence and logical consistency. Effective citations exhibit topological adaptation to the rhetorical structure of the host text, enhancing discourse cohesion and helping readers understand the cited work's contribution to the current research. Through discourse analysis, we define three integration models: (a) Bridging (connecting different arguments), (b) Supporting (providing evidential support), and (c) Contrasting (establishing theoretical distinctions). Each model requires specific syntactic framing strategies to ensure proper integration.

\subsubsection{Information Density}
Academic citations often contain rich information, which requires precise wording to condense the core content of the cited work. This condensation not only enhances information density but also enables readers to grasp the key points of the cited work without consulting the original source. High-quality citations efficiently convey information by skillfully integrating multiple layers of semantic content through precise vocabulary usage.

\subsection{CITEVAL Evaluation Framework}
Given the aforementioned characteristics, it is evident that traditional evaluation metrics based on word matching (such as ROUGE, BLEU, etc.) are inadequate for effectively assessing high-quality citation texts. This limitation manifests in two key aspects: First, these algorithms overly rely on lexical overlap statistics, overlooking the deeper semantic relationships within the text, such as the inability to recognize semantic enrichment brought by synonymous substitution and conceptual reorganization. Second, their narrow evaluation dimensions conflict with the multifaceted value orientation of academic citations, as they fail to assess the logical rigor of the argumentation or the innovativeness of academic perspectives.

In light of the limitations of traditional evaluation methods and inspired by long-text generation evaluation approaches, we have designed the CITEVAL multidimensional evaluation framework to assess the overall quality of generated citations from multiple perspectives, with large language models (LLMs) acting as the evaluators. Based on the four core dimensions of citation construction—target alignment, semantic fidelity, contextual integration, and information compression—we assign different weights to each dimension and calculate a composite average score. The specific evaluation aspects and prompt words can be found in the appendix.

\section{Experiment}
\subsection{Citation Article Recommendation Evaluation}
We evaluated our citation article recommendation method on four datasets, including ACL-200, FullTextPeerRead, RefSeer and arXiv, where two datasets, ACL-200 and FullTextPeerRead, are on the small side, while the RefSeer and arXiv data are on the large side. The experiments were evaluated using two metrics, MRR and R@K.

According to Table \ref{Recall Results}, our method Atten-CF, which integrates the two collaborative filtering algorithms, ScCF and CsCF, outperforms the HAtten model with a single encoder structure. This indicates that considering the two algorithms can improve our model. This improvement is particularly significant on the small datasets ACL-200 and FullTextPeerRead, compared to the large datasets. This is because the small datasets are more concise and the structure of the citation network is clearer, allowing the simple collaborative filtering approach to better capture the relationships between the documents. Conversely, as the size of the dataset increases, the citation network becomes more complex, and simple methods may not be sufficient to fully utilize complex citation relationships.

\begin{table*}[ht]
\centering
\renewcommand{\arraystretch}{1.25}
\begin{tabular}{l|ccccc}
\hline
\multicolumn{1}{c|}{Model} & R-1   & R-2   & R-L   & CITEVAL & Human \\ \hline
EXT-ORACLE                 & 22.21 & 4.96  & 16.84 & -       & -     \\ \hline
PTGEN                      & 24.6  & 6.16  & 19.19 & -       & -     \\
PTGEN-Cross                & 27.08 & 7.14  & 20.61 & -       & -     \\
BART-large                 & 29.62 & 9.86  & 24.51 & -       & -     \\ \hline
GPT-4 \cite{42}                     & -     & -     & -     & 82.59   & 84.87 \\
Deepseek-Chat-v3 \cite{48}          & -     & -     & -     & 82.91   & 80.96 \\
Qwen2.5-32B-Instruct \cite{49}      & -     & -     & -     & 72.11   & 76.53 \\ \hline
SciRGC                     & 37.54 & 15.63 & 28.74 & 84.22   & 82.04 \\ \hline
\end{tabular}
\caption{Comparison of automated assessment results between the SciRGC framework and baseline modeling}
\label{citation results}
\end{table*}

\begin{figure*}[ht]
\centering
\includegraphics[width=0.5\textwidth]{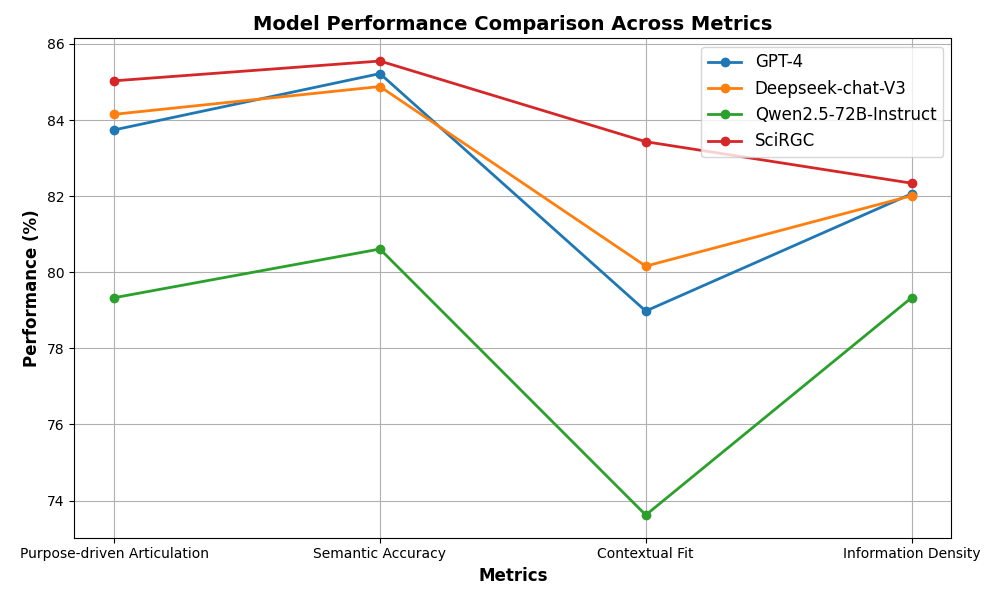}
\caption{Evaluation Results of Various Language Models Under the CITEVAL Metrics}
\label{fig5}
\end{figure*}

The evaluation results of the reranking model are shown in Table \ref{Reranking Results}. According to the table, the reordering model significantly outperforms the previous state-of-the-art model on all four datasets. This optimization not only indicates that the advantage of SciBERT's pre-training on a corpus of scientific papers is more conducive to the citation reordering task compared to general BERT models. It also indicates the introduction of citation function features enables the model to understand and utilize citation relationships between documents more effectively, which significantly improves the accuracy and efficiency of retrieval.

\subsection{Citation Sentence Generation Evaluation}

This paper evaluates the citation sentence generation method using the dataset provided by \cite{14}, and reports the ROUGE scores, CITEVAL scores, and human evaluation results in Table \ref{citation results}. The results show that the SciRGC method significantly outperforms the baseline models in terms of ROUGE scores, and performs comparably to larger models like GPT-4 and Deepseek on the CITEVAL metric.

\subsubsection{Traditional machine evaluation metrics}
This paper compares the performance of traditional evaluation methods and previous model architectures using the ROUGE metric. As shown in Table \ref{citation results}, extractive text summarization methods are not well-suited for citation generation tasks. This is because academic citations often involve novel expressions and require high-level abstraction or adaptation of the cited content, rather than simple text copying. In contrast, the SciRGC method demonstrates superior performance. This is not only due to its larger parameter scale but also because it optimizes sentence style through preference alignment techniques. Moreover, SciRGC integrates chain-of-thought reasoning to identify citation intent and key information, significantly enhancing text generation quality. As a result, it more accurately captures the original content and generates citation sentences that better align with the author's intent.

\subsubsection{CITEVAL evaluation metrics}
To address the evaluation challenges of large-scale language models, this study proposes and validates the CITEVAL multidimensional evaluation framework. As shown in Table \ref{citation results}, CITEVAL exhibits a strong positive correlation with human evaluation ($r = 0.82, p < 0.001$), demonstrating its effectiveness in assessing citation quality.  Notably, different dimensions show varying degrees of correlation with human ratings: semantic fidelity and target alignment have the greatest impact on human judgments, while information compression has the lowest correlation. Further analysis reveals that when citation length exceeds 25 words, information compression negatively correlates with readability, indicating the need to balance conciseness and fluency in the citation generation process.

As shown in Figure \ref{fig5}, different models exhibit distinct capability profiles. SciRGC demonstrates superior performance in structured metrics, with its target alignment ($85.03 \pm 1.2$) and contextual coherence ($83.43 \pm 0.9$) surpassing GPT-4 by 1.5\% and 5.6\%, respectively ($t$-test, $p < 0.05$).  GPT-4 achieves the best performance in semantic fidelity ($85.22$) and information compression ($82.06$), with a significantly lower term variation rate compared to Deepseek-V3. Leveraging an enhanced contextual modeling mechanism, Deepseek-V3 surpasses GPT-4 in contextual coherence ($80.16$).  The original Qwen2.5-32B model performs slightly worse, with a contextual coherence score of $73.62$—$9.81$ points lower than the best-performing model. Case analysis reveals that 32\% of its outputs exhibit logical inconsistencies (e.g., "The method in reference [12]... performs well" lacks supporting data).

\section{Conclusions}
This study proposes the SciRGC framework, specifically designed for citation retrieval and generation at specific positions in scientific articles. The framework integrates three modules: retrieval, intent classification, and generation, significantly enhancing the efficiency and quality of citation work.  In the citation retrieval module, the framework adopts a recall and re-ranking architecture, improving recall accuracy through citation networks and an Encoder architecture, and introduces citation intent features for re-ranking optimization. In the citation generation module, it leverages the advantages of large-scale language models, combining citation intent and retrieved article summaries. Using a chain-of-thought reasoning method, high-quality citation sentences are generated and aligned with human preferences. Additionally, based on the key features constructed from citations, the CITEVAL evaluation method is proposed.  Experimental results show that compared to baseline models, SciRGC achieves significant improvements in both citation retrieval accuracy and citation generation quality, while substantially reducing inference costs compared to models like GPT-4.

\appendix

\section{Citing Intent Categories}

In the citation intent recognition model, we mainly used three categories Background, Method and Comparative, which are defined as shown in Table \ref{intent Categories}
\begin{table}[ht]
\centering
\renewcommand{\arraystretch}{1}
\begin{tabular}{c|l}
\hline
\textbf{Categories} & \multicolumn{1}{c}{\textbf{Description}}                                                                                                                                                                                          \\ \hline
Background          & \begin{tabular}[c]{@{}l@{}}The citation states, mentions, or points to \\ the background information giving more context \\ about a problem, concept, approach, topic, \\ or importance of the problem in the field.\end{tabular} \\ \hline
Method              & \begin{tabular}[c]{@{}l@{}}Making use of a method, tool, approach or \\ dataset\end{tabular}                                                                                                                                      \\ \hline
Comparative         & \begin{tabular}[c]{@{}l@{}}Comparison of the paper's results/findings \\ with the results/findings of other work\end{tabular}                                                                                                     \\ \hline
\end{tabular}
\caption{The referential intent we use and its definition}
\label{intent Categories}
\end{table}

\begin{figure}[ht]
    \centering
    \includegraphics[width=0.5\columnwidth]{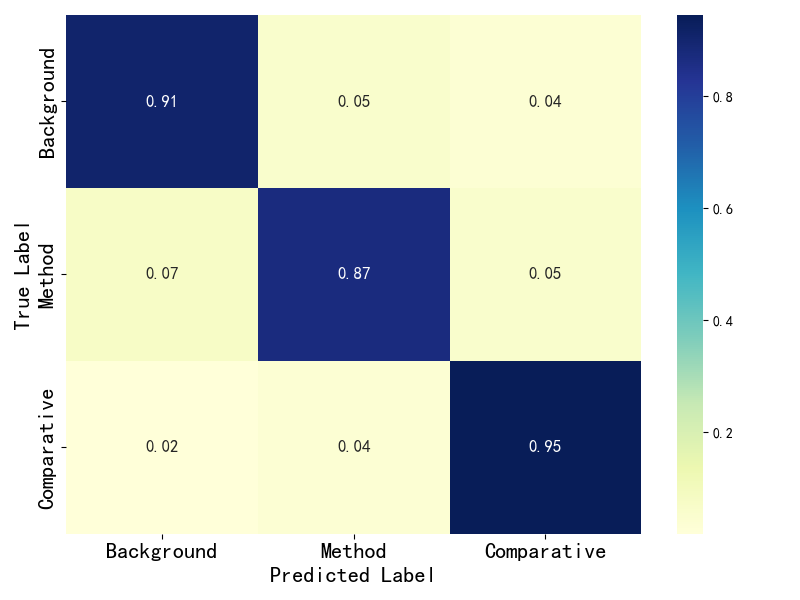} 
    \caption{Confusion matrix results by category}
    \label{fig 5}
\end{figure}

We further fine-tuned it by inserting a special marker [CLS] at the beginning of each citation sentence, which was then fed into a multilayer perceptron (MLP). Finally, a softmax function is used to predict the citation function of the citation sentence. We trained the model on the SciCite dataset and performed 10-fold cross-validation. The hidden layer size of the model MLP was set to 256 with a dropout of 0.1. The optimizer used was Adam \cite{26}, the learning rate was set to 1e-4, and the batch size was 64. Subsequently, in order to evaluate the robustness of the citation function classification model, we selected 400 citation sentences from the citation generation dataset \cite{3} and manually labeled their citation functions to form an independent external test set to validate the generalization performance of the model. We used accuracy, recall and F1 scores to assess the accuracy of the labeling. Figure \ref{fig 5} corresponds to each class confusion matrix. 

These results show that the model has high accuracy in the citation intent recognition task, especially in the "contrast" category, where it performs best and has the lowest misclassification rate. Overall, the experimental results validate the effectiveness of the model in this classification task, providing more accurate classification results, and its performance on external test sets also shows its good generalization performance.

\section{CITEVAL Scoring Guidelines}

\begin{itemize}
    \item \textbf{Purpose-driven Articulation (0.35)}
    \begin{description}
        \item[90-100:] Directly supports core arguments, forming indispensable logical closure.
        \item[80-89:] Effectively addresses main research questions, enhancing credibility.
        \item[70-79:] Provides relevant background without direct support.
        \item[60-69:] Partial relevance with limited argumentative value.
        \item[0-59:] Severe deviation or misleading content.
    \end{description}
    
    \item \textbf{Semantic Accuracy (0.25)}
    \begin{description}
        \item[90-100:] Fully preserves key elements with 100\% term accuracy.
        \item[80-89:] No core content distortion, reasonable simplification of minor parameters.
        \item[70-79:] Non-critical phrasing differences.
        \item[60-69:] Important concept/data inaccuracies.
        \item[0-59:] Substantive errors or fabricated content.
    \end{description}

    \item \textbf{Contextual Fit (0.25)}
    \begin{description}
        \item[90-100:] Seamlessly embedded, significantly enhances reading efficiency.
        \item[80-89:] Smooth transitions without comprehension barriers.
        \item[70-79:] Requires moderate cognitive adjustment.
        \item[60-69:] Causes localized logical disruptions.
        \item[0-59:] Severely compromises textual coherence.
    \end{description}

    \item \textbf{Information Density (0.20)}
    \begin{description}
        \item[90-100:] Pareto-optimal compression (balance of completeness and conciseness).
        \item[80-89:] Lossless compression of core information.
        \item[70-79:] Non-essential redundancy present.
        \item[60-69:] Critical omissions or oversimplification.
        \item[0-59:] Severe information density imbalance.
    \end{description}
\end{itemize}


\end{document}